\documentclass[12pt]{iopart}
\usepackage{iopams}  
\expandafter\let\csname equation*\endcsname\relax
\expandafter\let\csname endequation*\endcsname\relax

\usepackage{cite}
\usepackage{graphicx} 
\usepackage{dcolumn}
\usepackage{bm}
\usepackage{mathtools}

\usepackage{amsmath}
\usepackage{float}
\usepackage{subfig}
\usepackage{tikz}
\usepackage{pgffor} 
\usepackage{hyperref}

\begin{document}

\title[]{Reputation-based synergy and discounting mechanism promotes cooperation}

\author{Wenqiang Zhu$^1$$^,$$^2$, Xin Wang$^1$$^,$$^2$$^,$$^3$$^,$$^4$$^,$$^5$$^,$$^7$, Chaoqian Wang$^9$, Longzhao Liu$^1$$^,$$^2$$^,$$^3$$^,$$^4$$^,$$^5$$^,$$^7$$^,$$^*$, Hongwei Zheng$^1{^0}$$^,$$^*$ and Shaoting Tang$^1$$^,$$^2$$^,$$^3$$^,$$^4$$^,$$^5$$^,$$^6$$^,$$^7$$^,$$^8$}

\address{$^1$ Institute of Artificial Intelligence, Beihang University, Beijing, 100191, China}
\address{$^2$ Key laboratory of Mathematics, Informatics and Behavioral Semantics, Beihang University, Beijing 100191, China}
\address{$^3$ Zhongguancun Laboratory, Beijing 100094, China}
\address{$^4$ Beijing Advanced Innovation Center for Future Blockchain and Privacy Computing, Beihang University, Beijing 100191, China}
\address{$^5$ PengCheng Laboratory, Shenzhen 518055, China}
\address{$^6$ Institute of Medical Artificial Intelligence, Binzhou Medical University, Yantai 264003, China}
\address{$^7$ State Key Lab of Software Development Environment, Beihang University, Beijing 100191, China}
\address{$^8$ School of Mathematical Sciences, Dalian University of Technology, Dalian 116024, China}
\address{$^9$ Department of Computational and Data Sciences, George Mason University, Fairfax, VA 22030, USA}
\address{$^1{^0}$ Beijing Academy of Blockchain and Edge Computing, Beijing 100085, China}
\address{$^*$ Author to whom any correspondence should be addressed.}

\eads{\mailto{longzhao@buaa.edu.cn}, \mailto{hwzheng@pku.edu.cn}}

\begin{abstract}
A good group reputation often facilitates more efficient synergistic teamwork in production activities. Here we translate this simple motivation into a reputation-based synergy and discounting mechanism in the public goods game. Specifically, the reputation type of a group, either good or bad determined by a reputation threshold, modifies the nonlinear payoff structure described by a unified reputation impact factor. Results show that this reputation-based incentive mechanism could effectively promote cooperation compared with linear payoffs, despite the coexistence of synergy and discounting effects. Notably, the complicated interactions between reputation impact and reputation threshold result in a sharp phase transition from full cooperation to full defection. We also find that the presence of a few discounting groups could increase the average payoffs of cooperators, leading to an interesting phenomenon that when the reputation threshold is raised, the gap between the average payoffs of cooperators and defectors increases while the overall payoff decreases. Our work provides important insights into facilitating cooperation in social groups.
\end{abstract}

\noindent{\it Keywords\/}:{Public goods game, Evolution of cooperation, Reputation, Nonlinear payoffs}

\maketitle

\section{Introduction}
Cooperation is a pervasive behavior in nature and human society~\cite{1,2,3,4}, but understanding the emergence and maintenance of cooperation among selfish individuals in the context of Darwinian evolution remains a mystery~\cite{5,6}. Generally speaking, when there is a conflict between collective and individual interests, selfish individuals will always tend to pursue individual interests, thus resulting in the loss of collective interests, which is also known as social dilemma~\cite{7,8}. Over the decades, scholars in different fields have conducted in-depth research on this problem, in which evolutionary game theory~\cite{9,10,11,13} has been widely applied as a basic framework. In addition, many classical examples such as prisoner's dilemma game (PDG)~\cite{14,15,nowak1992evolutionary,16}, public goods game (PGG)~\cite{17,18,19,20} have been proposed to study different dilemmas.

In 2006, Nowak~\cite{21} outlined five mechanisms that promote the evolution of cooperation: kin selection, group selection, indirect reciprocity, direct reciprocity, and spatial reciprocity. Building on this foundation, numerous other mechanisms such as reputation~\cite{22,23,24,li2022n,xia2022costly}, rewards and punishments~\cite{25,26,szolnoki2011phase,szolnoki2011competition,27,28}, confidence~\cite{29,30,wang2023evolution,wang2023inertia,wang2023conflict,31}, aspiration~\cite{32,33}, and memory~\cite{34} have also been proven to be important for promoting cooperation. Within the domain of reputation-based mechanisms, the blueprints of how individual reputation is determined and how the corresponding social norms influence cooperation evolution have been roughly established~\cite{wei2023indirect}. Nowak~\cite{22} introduced the concept of ``image scoring" as a simplification of reputation. Furthermore, the allocation of reputation based on historical behavior has been extensively examined~\cite{35,36,37}. Duca and Nax~\cite{38} argued that allocating reputation at the group level might diminish its effectiveness in fostering cooperation. Ohtsuki and Iwasa~\cite{39} compiled the ``leading eight'' norms that can enhance cooperation. Along this line, more refined models with multifaceted dynamics involving real-world factors have also been explored. On the one hand, the reputation dynamics could trigger other complex behaviors, resulting in coupling mechanisms that have been shown to be effectively facilitate cooperation, such as  reputation-based partner choice~\cite{40,41}, interaction~\cite{42}, popularity~\cite{43}, migration~\cite{44} and others~\cite{45,46,47}, have also been demonstrated to effectively facilitate cooperation. On the other hand, reputation could directly affect the payoff structure of the game, which can be widely observed in production activities involving team collaboration~\cite{59,60}. For example, a good reputation can enhance team cohesion~\cite{61}. Team members have confidence in maintaining good cooperative relationships, bringing about more efficient synergistic teamwork and naturally more benefits~\cite{4}.

In fact, within diverse settings ranging from natural ecosystems to human societies, payoffs derived from group interactions often exhibit nonlinear characteristics~\cite{48,49,50,51}, most notably in the forms of synergy and discounting effects. For instance, yeast cells produce and secrete enzymes during foraging activities that decompose their environment, thereby supplying nutrients as public goods for the collective. While the initial cooperating cell is pivotal, the utility of additional enzymes diminishes as more cooperating cells join the ensemble. When cell density reaches saturation, further additions become redundant, illustrating a discounting effect~\cite{19,48}. For enzyme-facilitated reactions, the efficiency of subsequent enzymes can surpass linear projections as their numbers increase, demonstrating a synergistic effect~\cite{52}. Hauert~{\it et~al.}~\cite{48} utilized the notions of synergy and discounting to describe how payoffs accumulate in groups with multiple cooperators. Li~{\it et~al.}~\cite{53} explored the emergence and stability of cooperation under various conditions, including discounting, linear, and synergistic interactions within structured populations. Li and Wang~\cite{54} further examined the evolutionary dynamics of infinite structured population interactions and indicated that synergistic interactions within a group do not universally encourage cooperation. Jiang~{\it et~al.}~\cite{55} analyzed different evolutionary outcomes in nonlinear interactions based on global environmental fluctuations and local environmental feedback. Considering the coexistence of synergy and discounting in interactive groups, Zhou~{\it et~al.}~\cite{56} analyzed how periodic shifts between synergistic and discounting interactions affect evolution of cooperation on different complex networks. Zhou~{\it et~al.}~\cite{57} also developed a model for the co-evolution of synergy and discounting with individual strategies in well-mixed and structured populations, taking into account the positive and negative correlation of nonlinear interaction factors with the frequency of cooperators in a population, as well as local and global information, to further understand the evolution of cooperation. Quan~{\it et~al.}~\cite{58} introduced an economic scale threshold where the contribution of each additional cooperator to the group accrues in the form of a discount when the number of cooperators in the group is less than the threshold, and in the form of synergy when it is greater than the threshold. Their results show that nonlinear payoffs based on synergy and discounting promote cooperation more significantly than linear payoffs. 

Despite the progress, the effect of reputation on cooperative behavior in nonlinear games has yet to be studied. Here, we take reputation as a pre-judgment condition on the basis of nonlinear public goods games, which is specifically reflected in the fact that the calculation of game payoffs conducted by groups with higher reputation is synergistic, while that of groups with lower reputation is discounted. We preliminary show how reputation mechanism affects the evolution of cooperation when the effects of synergy and discounting coexist. Comparisons with the original linear payoffs revealed that the greater the effect of reputation on the nonlinear payoffs the more it facilitated the emergence of cooperation. The increase in the reputation impact factor and the enhancement factor will, to some extent, weaken the influence of the reputation threshold on the population. However, once the reputation threshold reaches a certain level, the population undergoes a phase transition and cooperation disappears. In addition, the presence of a small number of discounting groups can help the cooperator to improve its average payoff. 

\section{Model}
In our model, all players are distributed on a square lattice of size \( L \times L \) with periodic boundary conditions. Each player occupies a node and can interact with its four nearest neighbors, so that the size of each group is \(G = 5\) and each player participates in a total of five games. Denote \(S_i\) the strategy of player \(i\). At the initial moment, each player has the same probability of choosing either cooperation (\(S_i=C\)) or defection (\(S_i=D\)). In the original PGG, each cooperator contributes a cost \( c \), which is then multiplied by an enhancement factor \( r \) ($r>1$) and distributed equally among all players in the group. Defectors, on the other hand, make no contribution and generate no benefits. Within a group \( g \) of size \( G \) centered on player \( i \), let \( n_C \) represent the number of cooperators. The payoff for a player within this group can be expressed as follows:
\begin{equation}
    P_{i}^{g}
    =
    \begin{cases} 
    \displaystyle{\frac{r c n_{C}}{G}-c},  & \mbox{if $S_{i}(t)=C$,} \\[1em]
    \displaystyle{\frac{r c n_{C}}{G}}, & \mbox{if $S_{i}(t)=D$.}
    \end{cases}
\end{equation}

However, in some cases, the benefits contributed by each cooperator are usually different, showing non-linearity due to synergy and discounting in the investment. Hauert~{\it et~al.}~\cite{48} introduced the parameter \( \omega \) to describe this non-linearity. The first cooperator contributes a benefit \( rc \) (i.e., \( \omega^0 rc \)), the second cooperator contributes \( \omega^1 rc \), and so forth. The \( i^{\text{th}} \) cooperator contributes \( \omega^{i-1} rc \). When \( \omega > 1 \), the benefits increase, manifesting as a synergy; when \( \omega < 1 \), the benefits decrease, manifesting as a discounting; and when \( \omega = 1 \), it degenerates into the original public goods game. When there are no cooperators in the game group \( g \) ($n_C=0$), the payoff for player $i$ \( P_i^g=0 \). When there are cooperators in game group \( g \) ($n_C\geq 1$), the payoff for player \( i \) can be expressed as follows:
\begin{equation}\label{eq_pig_w}
    P_{i}^{g}
    =
    \begin{cases} 
    \displaystyle{\frac{r c}{G}\left(1+\omega+\omega^{2}+\cdots+\omega^{n_{C}-1}\right)-c},  & \mbox{if $S_{i}(t)=C$,} \\[1em]
    \displaystyle{\frac{r c}{G}\left(1+\omega+\omega^{2}+\cdots+\omega^{n_{C}-1}\right)}, & \mbox{if $S_{i}(t)=D$.}
    \end{cases}
\end{equation}

Player \( i \) participates in a total of \(G = 5\) games, and the total payoff \( \pi_i \) is
\begin{equation}
   \pi_i=\sum_{g\in\Omega_i}P_i^g,
\end{equation}
where \( \Omega_i \) denotes the set of PGG groups in which agent \(i\) participates.

Let the reputation of player \( i \) at time \( t \) be \( R_i(t) \). We set \( R_i(t) \in [0, 100] \), considering the non-negativity and finiteness of reputation. If the strategy of player \( i \) at time \( t \) is cooperation, then his reputation increases by 1; otherwise, it decreases by 1. This categorization of reputation is similar to the ``image scoring'' defined by Nowak and Sigmund~\cite{22}, expressed as:
\begin{equation}\label{eq_Ri}
    R_i(t)
    =
    \begin{cases} 
    \displaystyle{R_i(t-1)+1},  & \mbox{if $S_{i}(t)=C$,} \\
    \displaystyle{R_i(t-1)-1}, & \mbox{if $S_{i}(t)=D$.}
    \end{cases}
\end{equation}

The reputation of the group centered on player \(i\), reflecting the level of goodness or badness of the environment, is determined by the reputation of all members within the group. Let \(\bar{R}_i(t)\) denote the reputation of the group centered on player \(i\), defined as:
\begin{equation}
\bar{R}_{i}(t)=\frac{\sum_{j\in\Omega_{i}}R_{j}(t)}{G}.
\end{equation}
The reputation of the group determines the accumulation or loss of gains within the group. A unified reputation impact factor, \(\delta\) (\(\delta \geq 0\)), is introduced to determine the non-linearity parameter \(\omega\), creating a bridge from reputation to gains. Specifically, if the reputation of the group is equal or greater than a reputation threshold \(A\), it is considered a high-reputation type group, rendering the gains exhibiting a synergistic effect. Conversely, if the group reputation is below this threshold, the group is considered a low-reputation type and the gains manifest a discount effect. Mathematically, it is defined as:
\begin{equation}\label{eq_w}
    \omega
    =
    \begin{cases} 
    \displaystyle{1+\delta},  & \mbox{if $\bar{R}_i(t)\geq A$,} \\
    \displaystyle{1-\delta}, & \mbox{if $\bar{R}_i(t)< A$.}
    \end{cases}
\end{equation}

The strategy update process employs asynchronous Monte Carlo simulations (MCS), ensuring that each player has, on average, one opportunity to be selected for update during a full MC step. A randomly selected player \(i\) chooses another player \(j\) from its neighbors, and pairwise comparison is conducted according to Fermi probability~\cite{16}:
\begin{equation}
    \Gamma_{(S_i\leftarrow S_j)}=\frac{1}{1+\exp[-(\pi_j-\pi_i)/K]},
\end{equation}
where \( K \) is the amplitude of noise, representing the degree of irrationality in players' decision-making. As \( K \) approaches zero, player \( i \) deterministically imitates the strategy of player \( j \) if \( j \) achieves a higher payoff. Conversely, as \( K \) tends toward infinity, the imitation of player \( j \) by player \( i \) becomes entirely random, resulting in a neutral drift within the population.

We focus on the fraction of cooperators within the population, noting the total number of cooperators in the population as \(N_C\), the fraction of cooperators in the population is denoted as \(\rho_C = N_C/L^2\). The final results are obtained by averaging over $1\times 10^1$ independent experiments, each of which runs for \( 2 \times 10^4 \) MCS, with the last \( 2 \times 10^3 \) steps being averaged.

\section{Results and analysis}
In all simulations conducted in this paper, we set \( L = 100 \), \( c = 1 \), and \( K = 0.5 \). To mitigate the impact of network size, we also run simulations on larger grids (e.g., $L\geq400$) and obtain qualitatively similar results. 

\begin{figure}
    \centering
    \includegraphics[width=8.3cm]{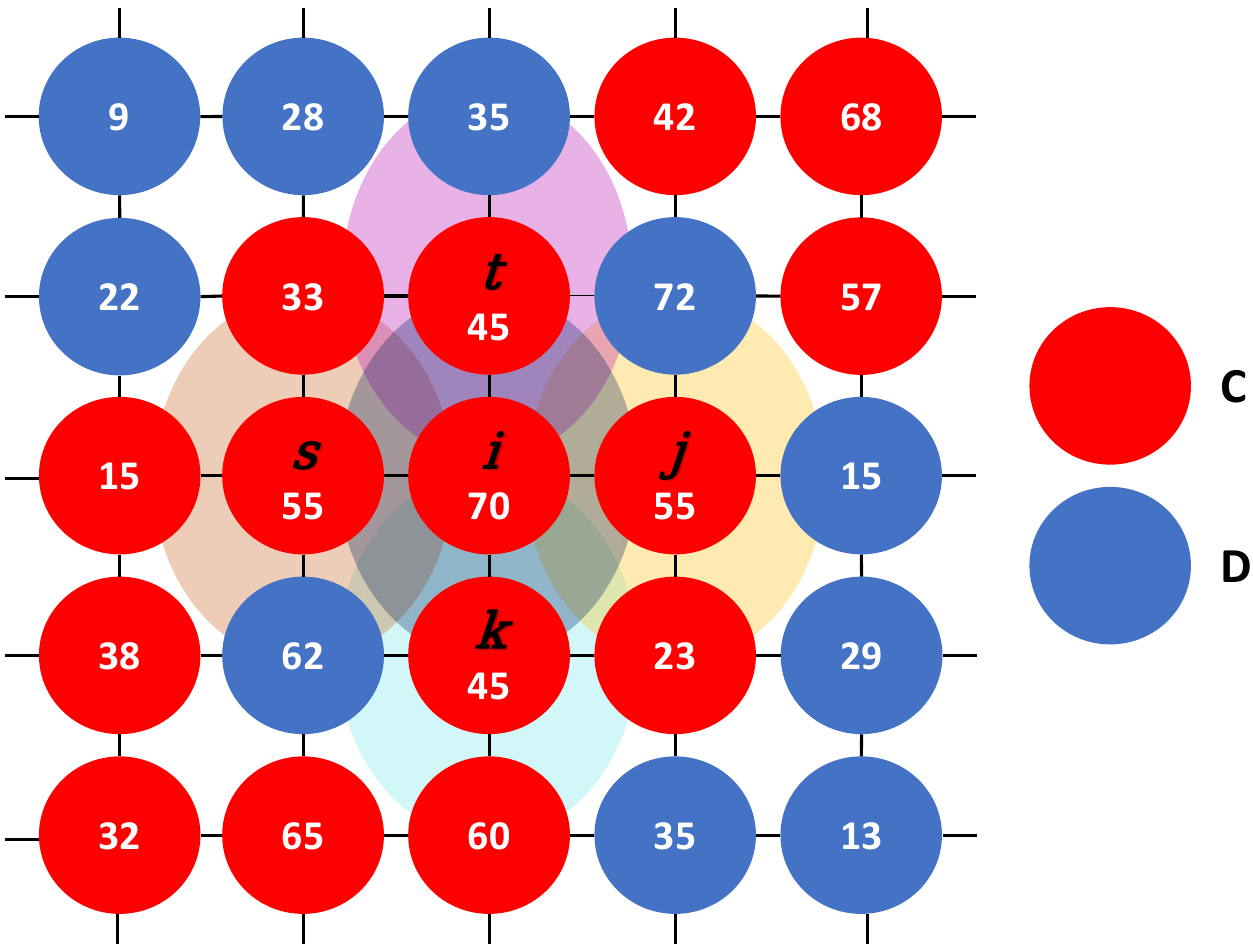}
    \caption{An example of payoff calculation based on the model. Here, the red spheres represent cooperators, while the blue spheres indicate defectors. The number within the spheres signify the reputation value of the players. Interactions between players are represented by the connecting lines.}\label{fig_diagram}
\end{figure}

We first provide an example of how the reputation-based synergy and discounting mechanism affects the payoff accumulation when \( A = 50 \), \( r = 4 \), and \( \delta = 0.1 \) in Figure~\ref{fig_diagram}. The central individual \( i \) participates in $G=5$ public goods games with groups centered on itself and its neighbors \( j, k, s, t \). In the group centered on player \( i \), \( \bar{R}_i = (70+45+55+45+55)/5=54 > 50 \). Therefore, it is a high-reputation group, and the gains are calculated to reflect the synergistic effect. According to Eq.~(\ref{eq_pig_w}) and Eq.~(\ref{eq_w}), player \( i \) obtains a benefit of 3.88408. In the groups centered on players \( j \) and \( t \), \( \bar{R}_j = (55+72+15+23+70)/5=47 < 50 \) and \( \bar{R}_t =(45+35+72+70+33)/5= 51 > 50 \). Although both groups have the same number of cooperators, due to the difference in group reputation, the gains that can be shared by the players in the two groups differ. Player \( i \) can obtain 1.648 in the group centered on \( t \), while only getting 1.168 in the group centered on \( j \). Similarly, the gains that player \( i \) can obtain in the groups centered on players \( k \) and \( s \) can be calculated to be 2.7128 and 1.7512, respectively. Therefore, in this round, player \( i \) can obtain a total benefit of 11.16408, an increase of 0.96408 compared to the original public goods game benefit of 10.2.

\begin{figure}[htbp]
    \centering
    \includegraphics[width=8.3cm]{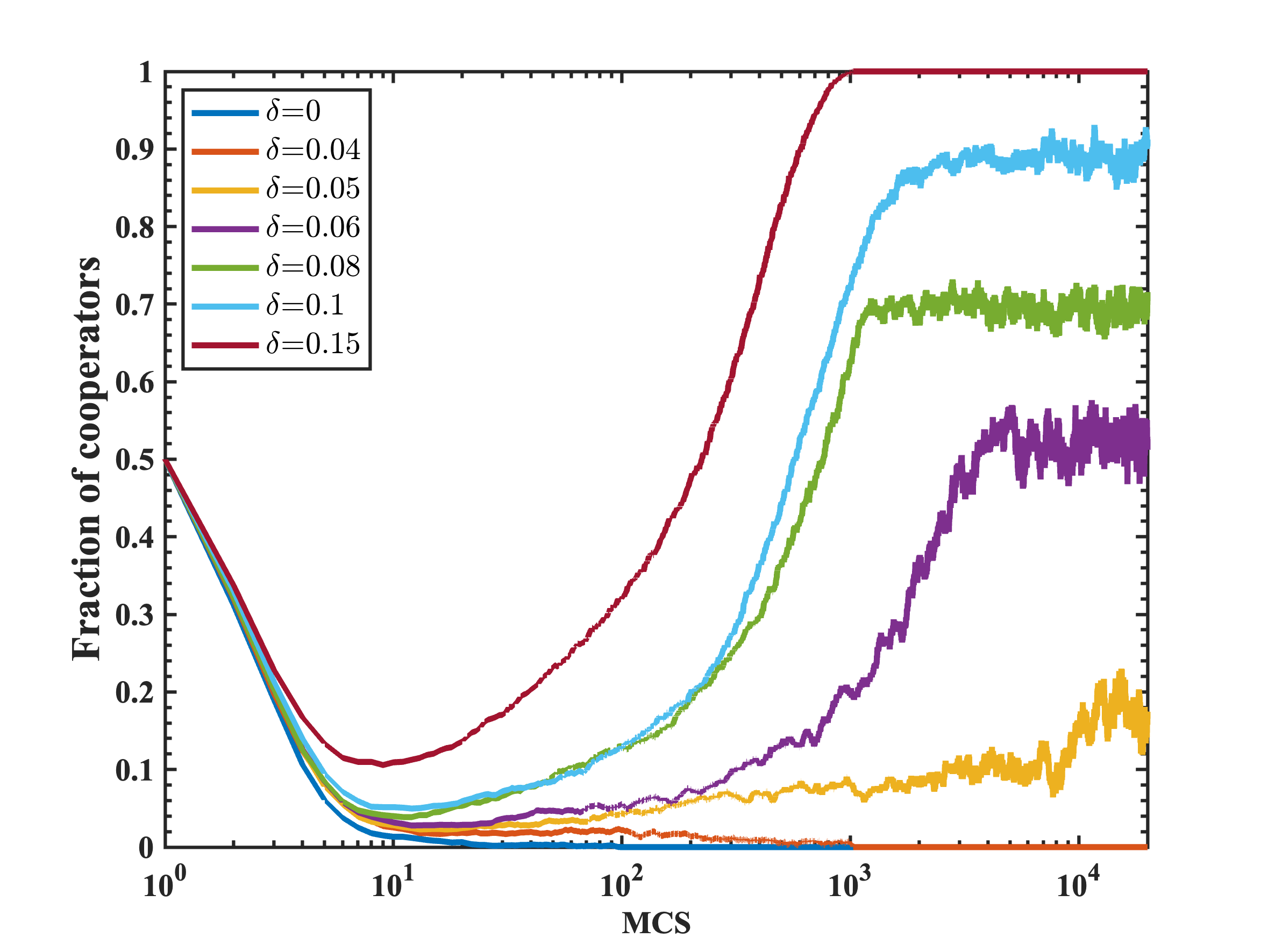}
    \caption{Evolution of the fraction of cooperators when \( r=3.5 \), \( A=50 \), and \( \delta \in \{0, 0.04, 0.05, 0.06, 0.08, 0.1, 0.15\} \).}\label{fig_timeevolution}
\end{figure}
Figure~\ref{fig_timeevolution} shows the evolution of the fraction of cooperators over time for different reputation impact factors $\delta$ when \( r = 3.5 \) and \( A = 50 \). When \( \delta = 0 \), the model reverts to the original public goods game, and cooperators eventually extinct. When \( \delta > 0 \), the effects of synergy and discounting are manifested in the game. When \( \delta = 0.04 \) is smaller, there is still no cooperator emergence as time evolves, which only prolongs the presence of cooperators compared to the original case. When \( \delta = 0.05 \), the fraction of cooperators first decreases and then increases over time, eventually stabilizing, allowing for the coexistence of cooperators and defectors in the population. Furthermore, as \( \delta \) continues to increase, the stable fraction of cooperators also improves, and when \( \delta = 0.15 \), only cooperators remain in the population. Thus, for a given enhancement factor, cooperation can only arise when the gains brought by reputation reach a certain level. And the greater the fluctuation in gains, the more favorable it is for cooperation.

\begin{figure}[htbp]
	\centering    
	\subfloat[] 
	{
		\begin{minipage}[t]{0.45\textwidth}
			\centering          
			\includegraphics[width=1\textwidth]{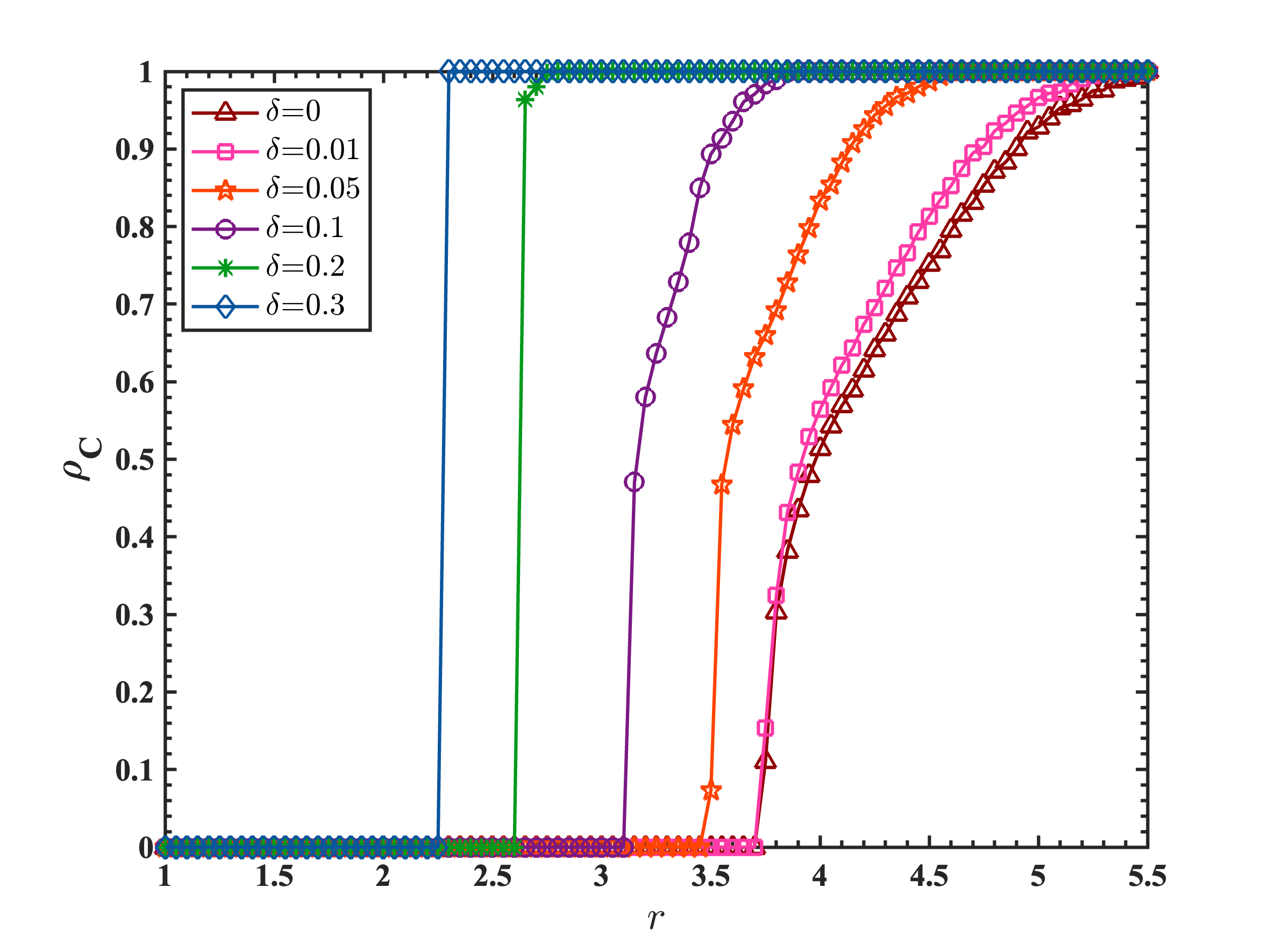}   
		\end{minipage}%
	}
	\subfloat[] 
	{
		\begin{minipage}[t]{0.45\textwidth}
			\centering      
			\includegraphics[width=1\textwidth]{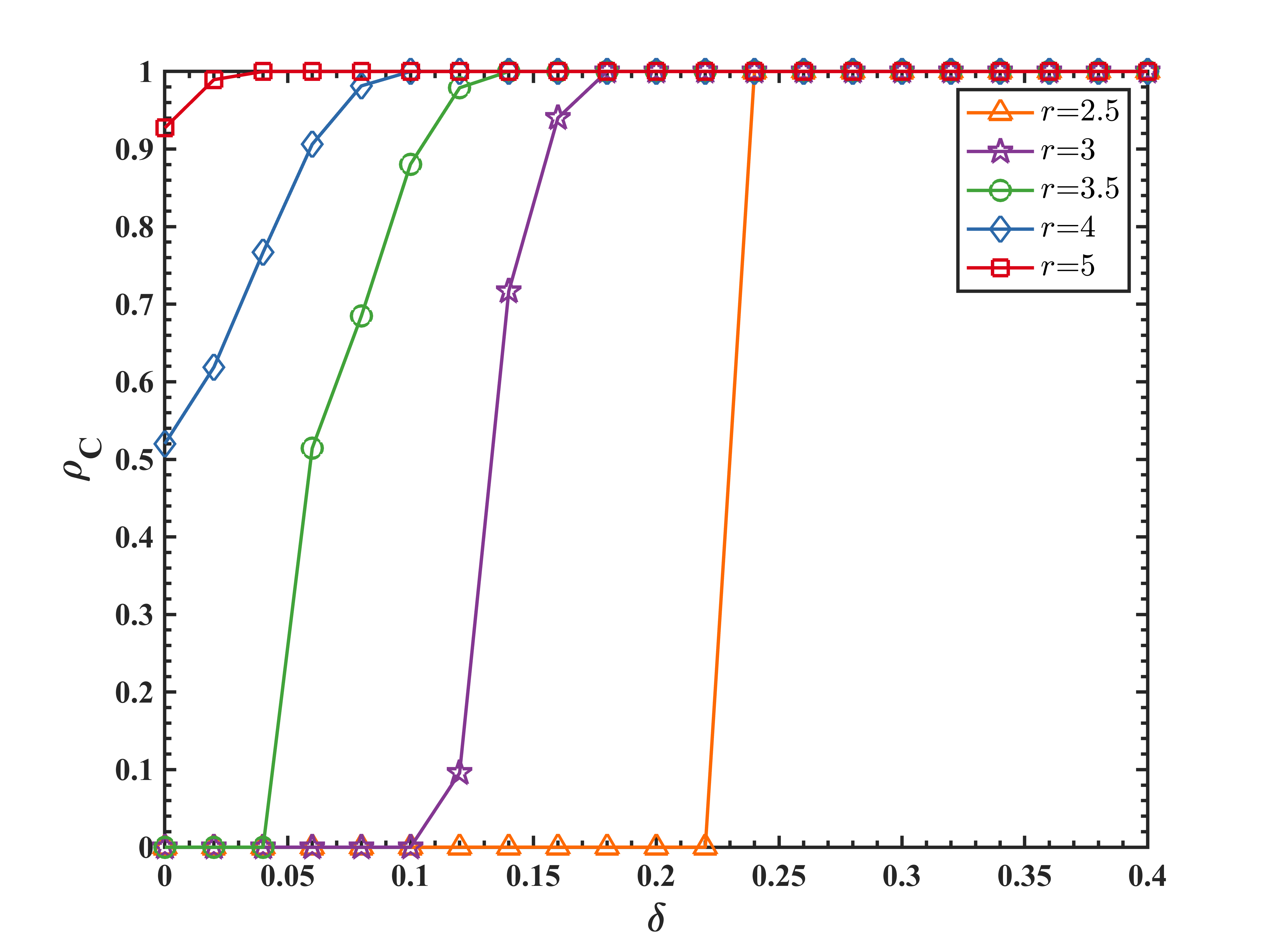}   
		\end{minipage}
	}
	\caption{(a) The evolution results of the fraction of cooperators $\rho_C$ with respect to $r$ when $\delta\in\{0,0.01,0.05,0.1,0.2,0.3\}$. (b) The evolution results of the fraction of cooperators $\rho_C$ with respect to $\delta$ when $r\in\{2.5,3,3.5,4,5\}$. The fixed parameter is $A=50$.}\label{fig_1D}
\end{figure}

Then, Figure~\ref{fig_1D} presents the evolutionary outcomes of the fraction of cooperators in terms of the enhancement factor \( r \) and the reputation impact factor \( \delta \) when \( A=50 \). As shown in Figure~\ref{fig_1D}(a), for \( \delta=0 \), cooperators emerge when \( r > 3.74 \) and defectors disappear when \( r > 5.49 \). In the region where both types coexist, the fraction of cooperators increases with \( r \)~\cite{26}. For a small \( \delta = 0.01 \), the \( r \) value at which cooperators appear is not very different from \( \delta = 0 \), but defectors disappear at a lower \( r \) value. Moreover, when cooperator and defector coexist, the fraction of cooperators is slightly higher for \( \delta = 0.01 \) than for \( \delta = 0 \). Therefore, even a small reputation impact factor can promote cooperation within certain \( r \) value ranges. As \( \delta \) continues to increase, the \( r \) values corresponding to the appearance of cooperators and the disappearance of defectors both decrease. Furthermore, the length of the \( r \) value interval where both types coexist also decreases. This suggests that an increasing reputation impact factor speeds up the phase transition from complete defection to complete cooperation, making it difficult for cooperators and defectors to coexist when the reputation impact factor reaches a certain level. In both Figure~\ref{fig_1D}(a) and Figure~\ref{fig_1D}(b), it can be observed that the fraction of cooperators increases with \( \delta \) for different \( r \) values. In Figure~\ref{fig_1D}(b), when \( r \) is large, even a small \( \delta \) can lead the population to a fully cooperative state. As \( r \) decreases, the \( \delta \) required to reach full cooperation increases. Furthermore, the length of the \( \delta \) interval where cooperators and defectors coexist first increases and then decreases. Specifically, when \( r = 2.5 \), the population undergoes a phase transition from complete defection to complete cooperation at larger \( \delta \). Therefore, when the enhancement factor is small, it is difficult to induce cooperation within the population, yet once the reputation impact factor reaches a certain value, the population quickly becomes fully cooperative. When the enhancement factor is large, cooperators easily survive, and considering reputation will also quickly lead the population to a fully cooperative state. When the enhancement factor is moderate, an increasing reputation impact factor can gradually help cooperators establish a greater advantage, allowing both cooperators and defectors to coexist within a larger range.

\begin{figure}[ht]
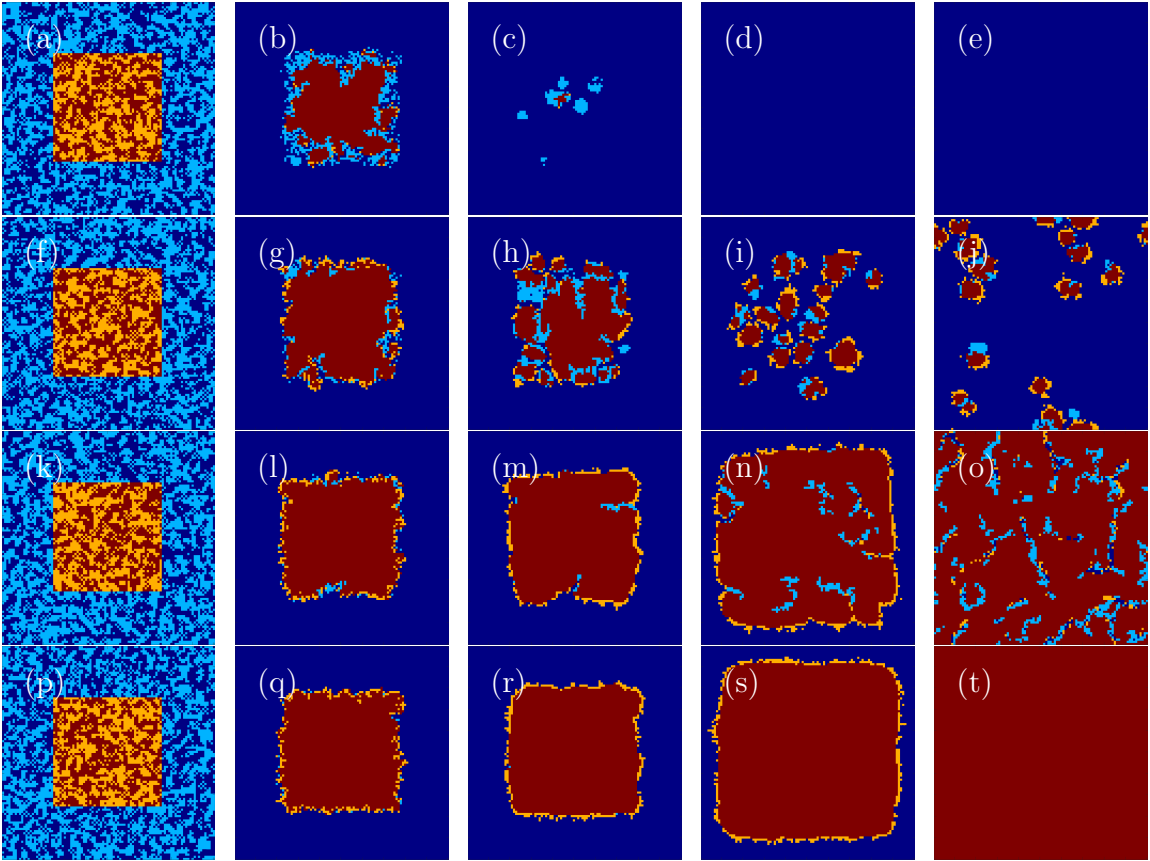

    \centering
    \begin{minipage}{1\textwidth}
        \centering
        \foreach \x/\y in {1/a, 2/b, 3/c, 4/d, 5/e} {
            \begin{tikzpicture}
                \node[anchor=north west,inner sep=0] at (0,0) {\includegraphics[width=0.18\textwidth,height=0.18\textwidth]{fig4-1-\x.png}};
                \node[anchor=north west, xshift=1.5mm, yshift=-1.5mm,text=white] at (0,0) {(\y)};
            \end{tikzpicture}
        } \hfill \\ 
        \foreach \x/\y in {1/f, 2/g, 3/h, 4/i, 5/j} {
            \begin{tikzpicture}
                \node[anchor=north west,inner sep=0] at (0,0) {\includegraphics[width=0.18\textwidth,height=0.18\textwidth]{fig4-2-\x.png}};
                \node[anchor=north west, xshift=1.5mm, yshift=-1.5mm,text=white] at (0,0) {(\y)};
            \end{tikzpicture}
        } \hfill \\ 
        \foreach \x/\y in {1/k, 2/l, 3/m, 4/n, 5/o} {
            \begin{tikzpicture}
                \node[anchor=north west,inner sep=0] at (0,0) {\includegraphics[width=0.18\textwidth,height=0.18\textwidth]{fig4-3-\x.png}};
                \node[anchor=north west, xshift=1.5mm, yshift=-1.5mm,text=white] at (0,0) {(\y)};
            \end{tikzpicture}
        } \hfill \\ 
        \foreach \x/\y in {1/p, 2/q, 3/r, 4/s, 5/t} {
            \begin{tikzpicture}
                \node[anchor=north west,inner sep=0] at (0,0) {\includegraphics[width=0.18\textwidth,height=0.18\textwidth]{fig4-4-\x.png}};
                \node[anchor=north west, xshift=1.5mm, yshift=-1.5mm,text=white] at (0,0) {(\y)};
            \end{tikzpicture}
        } 
    \end{minipage}
    \caption{Evolutionary snapshots starting from the same initial configuration and different reputation impact factor \( \delta \). Initially, the reputation of all players follows a uniform distribution in the range \([0,100]\), with cooperators concentrated in a central region of the population. For ease of distinction, red represents high-reputation group centered on cooperator (HC), orange represents low-reputation group centered on cooperator (LC), light blue represents high-reputation group centered on defector (HD), and dark blue represents low-reputation group centered on defector (LD). Each row represents a complete evolutionary process under different reputation impact factors. From left to right, the MCS are \(1, 50, 200, 1000, 20000\), respectively. From top to bottom, the reputation impact factors \( \delta \) are \(0, 0.05, 0.1, 0.15\), respectively. The fixed parameters are \( r=3.5 \) and \( A=50 \).}
    \label{fig_snap}
\end{figure}

To better understand the influence of different reputation impact factors on the evolution of cooperation, Figure~\ref{fig_snap} presents snapshots of the evolutionary trajectory starting from a typical distribution when \( r=3.5 \) and \( A=50 \) for several distinct \( \delta \) values. As evidenced by Figure~\ref{fig_snap}, the population eventually evolves into distinct spatial patterns depending on the value of \( \delta \). For \( \delta = 0 \), which represents the original public goods game, defector clusters invade cooperator clusters due to a clear advantage in gain when the reputation mechanism is inactive, as shown in Figure~\ref{fig_snap}(b). In Figure~\ref{fig_snap}(c), only a few cooperators remain, and as time further evolves, cooperators extinct, leaving only LD in the population. At \( \delta = 0.05 \), the reputation mechanism comes into play, introducing uncertainty in the gain differences between cooperators and defectors. Figure~\ref{fig_snap}(g) shows mutual invasions between cooperator and defector clusters, but Figure~\ref{fig_snap}(h) reveals that defectors invade cooperator clusters at a faster rate, with the original HC clusters largely taken over by HD. As time further evolves, in Figure~\ref{fig_snap}(i), HC clusters break down into smaller clusters, surrounded by a minority of LCs and HDs. By the time the population stabilizes, as shown in Figure~\ref{fig_snap}(j), all four types coexist, with a large number of LDs, and roughly equal numbers and distributions of LCs and HDs. For \( \delta = 0.1 \), there is the mutual invasion of cooperators and defectors in Figure~\ref{fig_snap}(l) as well, but due to the widening of the gain gap between the high-reputation groups and the low-reputation groups, it can be found from Figure~\ref{fig_snap}(m) that the cooperators invade the defector clusters at a faster rate, and HD shows a ``banded distribution'' in Figure~\ref{fig_snap}(n). With further evolution over time, when the population reaches stability, in Figure~\ref{fig_snap}(o), the four categories coexist in the population, at which point the cooperators prevail and the reputation in the population is generally high, with only a very small number of LCs and LDs. As \( \delta \) continues to increase to \( \delta = 0.15 \), Figure~\ref{fig_snap}(q) reveals that defectors no longer invade cooperator clusters. Furthermore, Figures~\ref{fig_snap}(r) and Figures~\ref{fig_snap}(s) show that HC and LD clusters are surrounded by LCs, with almost no HD present. As time evolves, the cooperator clusters continue to expand, leading the population into a fully cooperative state, as shown in Figure~\ref{fig_snap}(t). Thus, the mechanism can significantly enhance spatial reciprocity.

\begin{figure}[htbp]
	\centering    
	\subfloat[] 
	{
		\begin{minipage}[t]{0.45\textwidth}
			\centering          
			\includegraphics[width=1\textwidth]{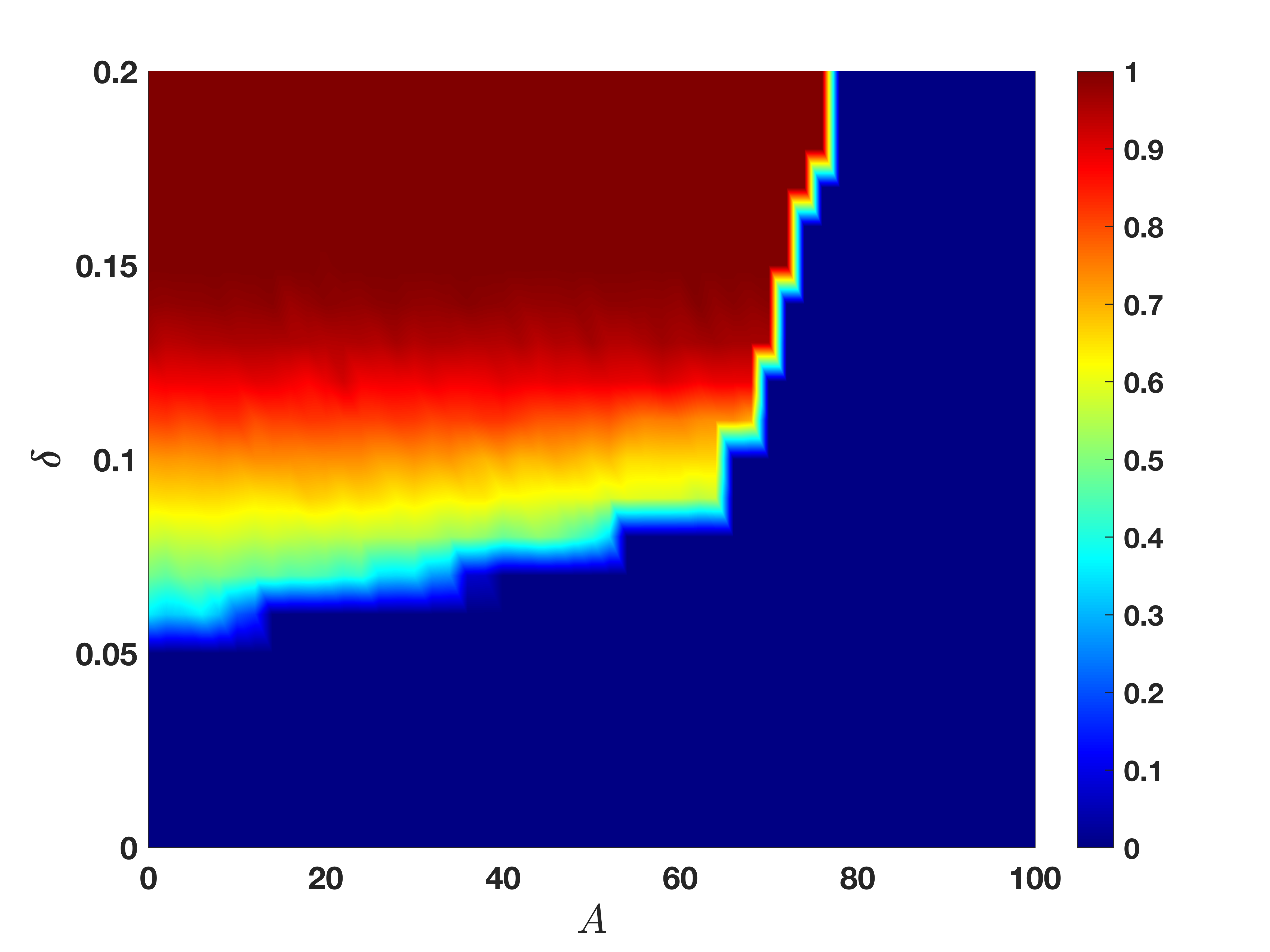}   
		\end{minipage}%
	}
	\subfloat[] 
	{
		\begin{minipage}[t]{0.45\textwidth}
			\centering      
			\includegraphics[width=1\textwidth]{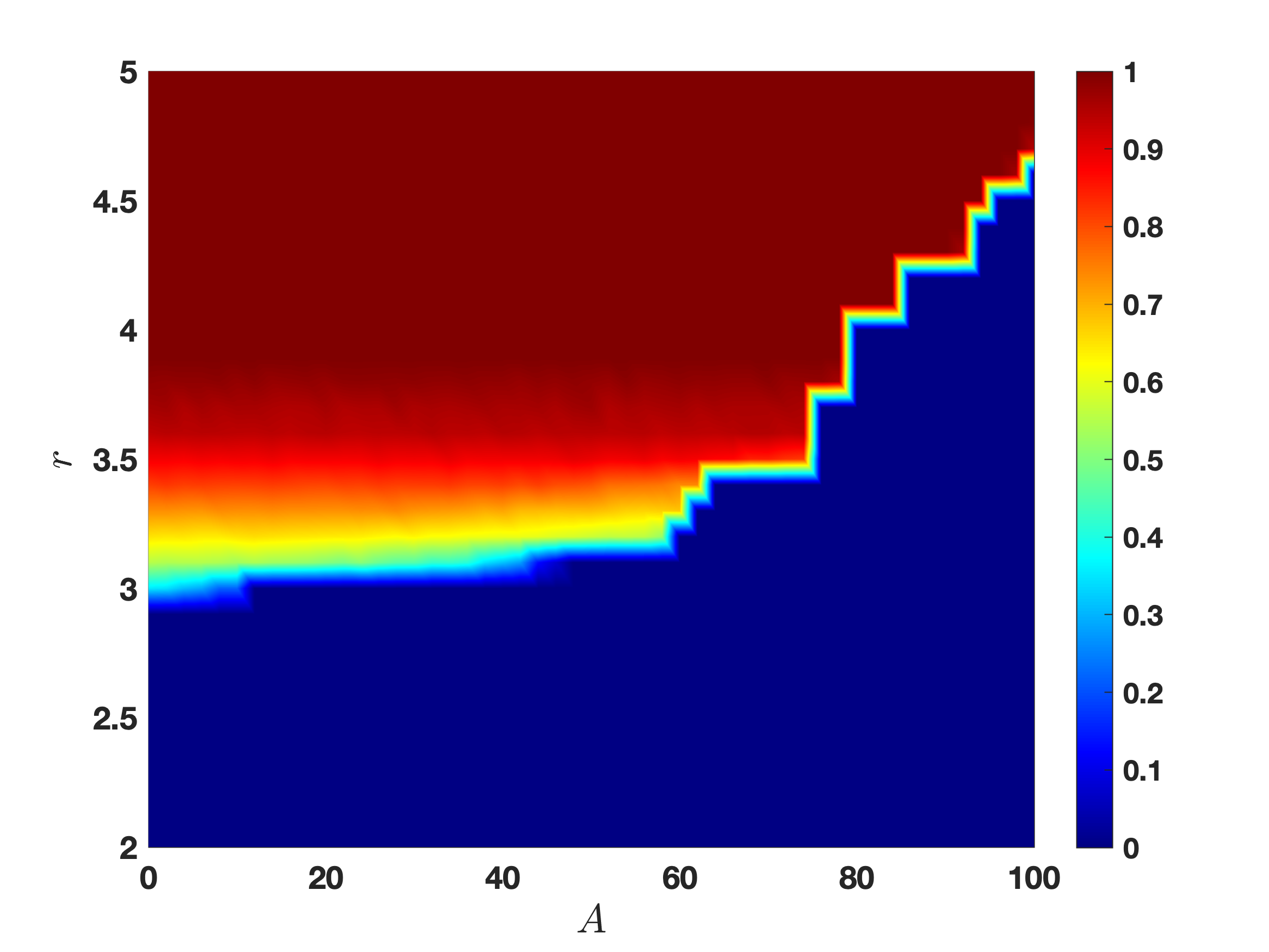}   
		\end{minipage}
	}
        \\
        \subfloat[] 
	{
		\begin{minipage}[t]{0.45\textwidth}
			\centering          
			\includegraphics[width=1\textwidth]{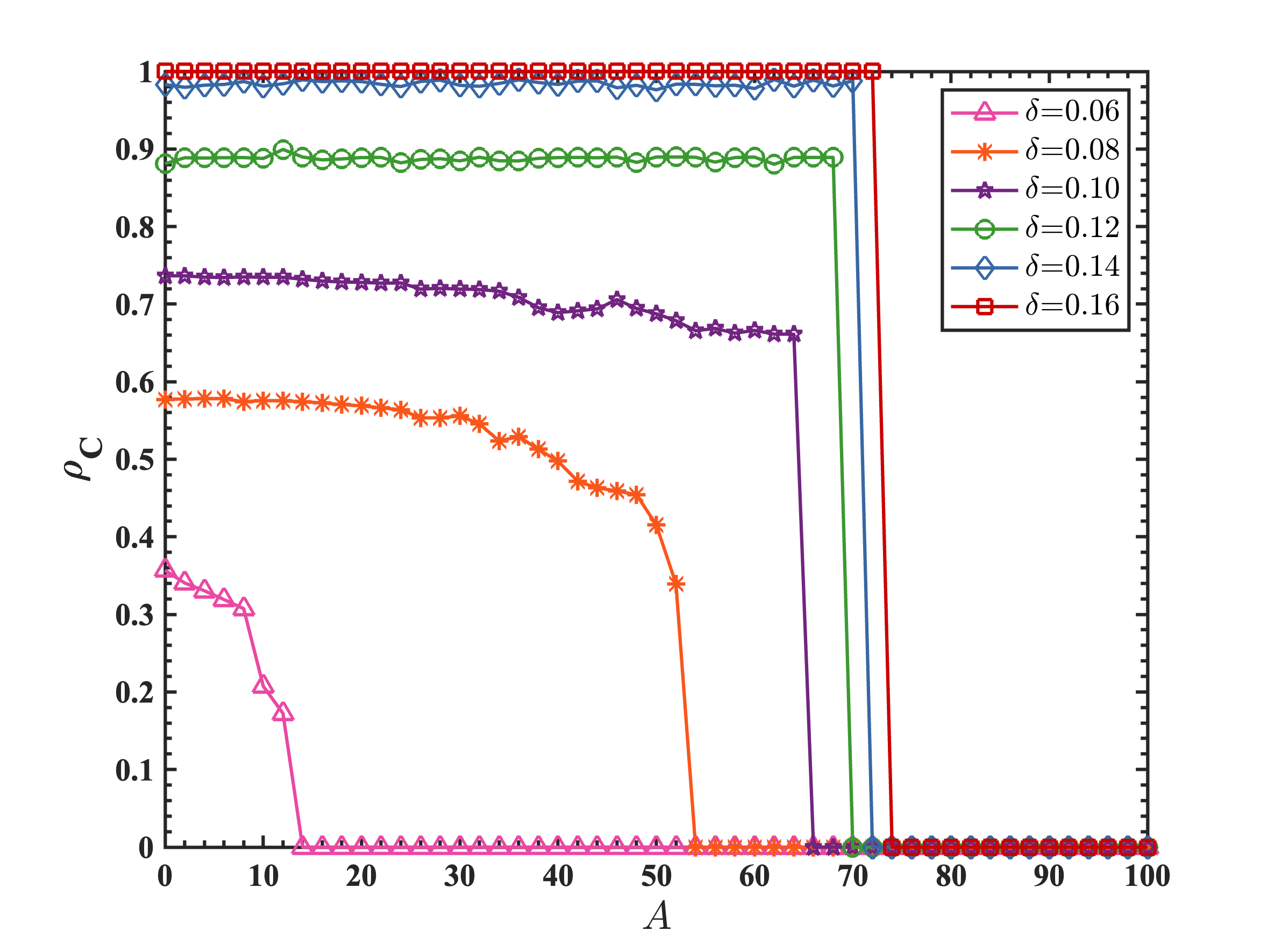}   
		\end{minipage}%
	}
	\subfloat[] 
	{
		\begin{minipage}[t]{0.45\textwidth}
			\centering      
			\includegraphics[width=1\textwidth]{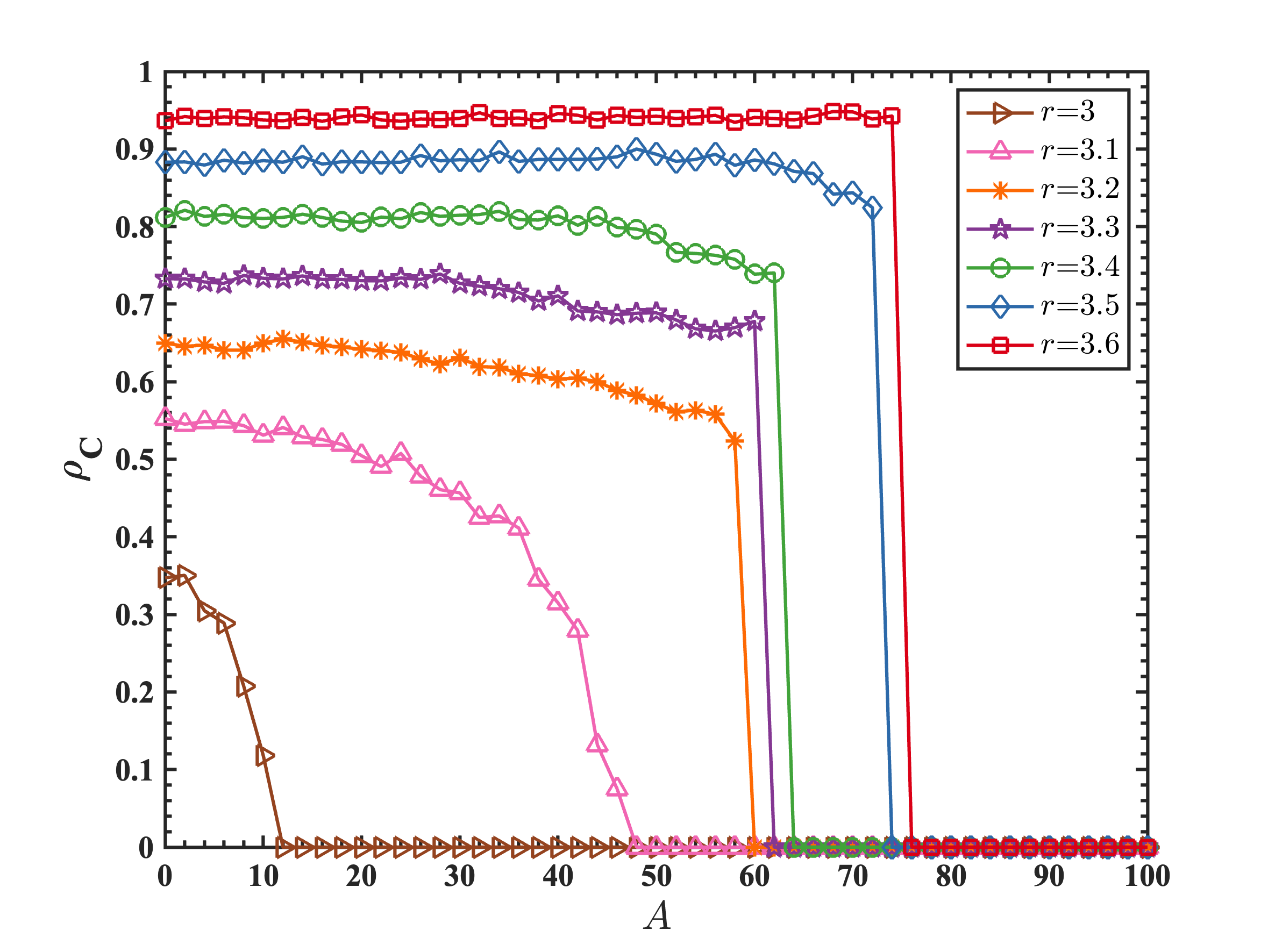}   
		\end{minipage}
	}
	\caption{Heat map of the fraction of cooperators with respect to the $A$-$\delta$ parameter plane (a) and $A$-$r$ parameter plane (b). Panels (c) and (d) show the evolution of the fraction of cooperators with respect to $A$ when some specific $\delta$ and specific $r$ are taken in panels (a) and (b), respectively. The fixed parameters are: (a,c) $r=3.3$ and (b,d) $\delta=0.1$.}\label{fig_threshold}
\end{figure}

The reputation threshold reflects how difficult it is for the group to become a high-reputation group, as well as the feedback from the game environment on the reputation of the players. If the gaming environment is lenient with respect to reputation, the number of high-reputation groups in the population will increase, and vice versa. Figure~\ref{fig_threshold} analyze the effect of reputation threshold on the evolution of cooperation under different reputation impact factors and enhancement factors. It is found that these two parameters show a similar pattern with the change of reputation threshold.

In Figure~\ref{fig_threshold}(a) and Figure~\ref{fig_threshold}(b), as \( A \) increases, the required \( \delta \)(or \(r) \) to induce cooperation gradually increases, meaning that a larger reputation impact factor (or enhancement factor) is needed to generate cooperation under higher reputation threshold conditions. Moreover, the length of the \( \delta \) (or \(r \)) interval where cooperators and defectors coexist is gradually decreasing as \( A \) increases. It is worth noting that within the parameter region of cooperator presence, the effect of changes in reputation thresholds on the fraction of cooperators shows two trends with different reputation impact factors (or enhancement factors). In Figure~\ref{fig_threshold}(c), when $\delta$ is small, the fraction of cooperators gradually decreases to zero as the reputation threshold increases, indicating that a lower reputation threshold is more favorable for cooperation. For a large $\delta$, however, changes in the reputation threshold over a range of values have little effect on the fraction of cooperators. But once the reputation threshold reaches a certain level, the fraction of cooperators suddenly drops to zero (although it remains high at slightly lower reputation thresholds). In this case, within a certain range of reputation thresholds, any reputation threshold has the same effect on promoting cooperation. The same phenomenon is also reflected in the evolutionary results seen in Figure~\ref{fig_threshold}(d) for different enhancement factors with respect to reputation threshold. This is because for larger $\delta$ and $r$, the synergistic group can generate more payoffs, while the payoffs of the discount group deteriorate. Spatial reciprocity allows cooperators to form a cluster~\cite{nowak1992evolutionary}, and a cluster of cooperators can easily become a synergistic group by accumulating reputation when the reputation threshold varies in a low range. Once a high-reputation cluster of cooperators reaches a certain size, it is difficult for defectors to invade. However, when the reputation threshold reaches a certain level, it takes a long process for a cluster of cooperators to gain synergistic benefits, in which they will be invaded by betrayers until they die out. That is, a certain range of reputation thresholds are allowed to vary without affecting the overall fraction of cooperators when the payoffs are high. Therefore, the increase in the reputation impact factor and the enhancement factor will, to some extent, weaken the influence of the reputation threshold on the population.

  \begin{figure}[htbp]
	\centering    
	\subfloat[] 
	{
		\begin{minipage}[t]{0.3\textwidth}
			\centering          
			\includegraphics[width=1\textwidth]{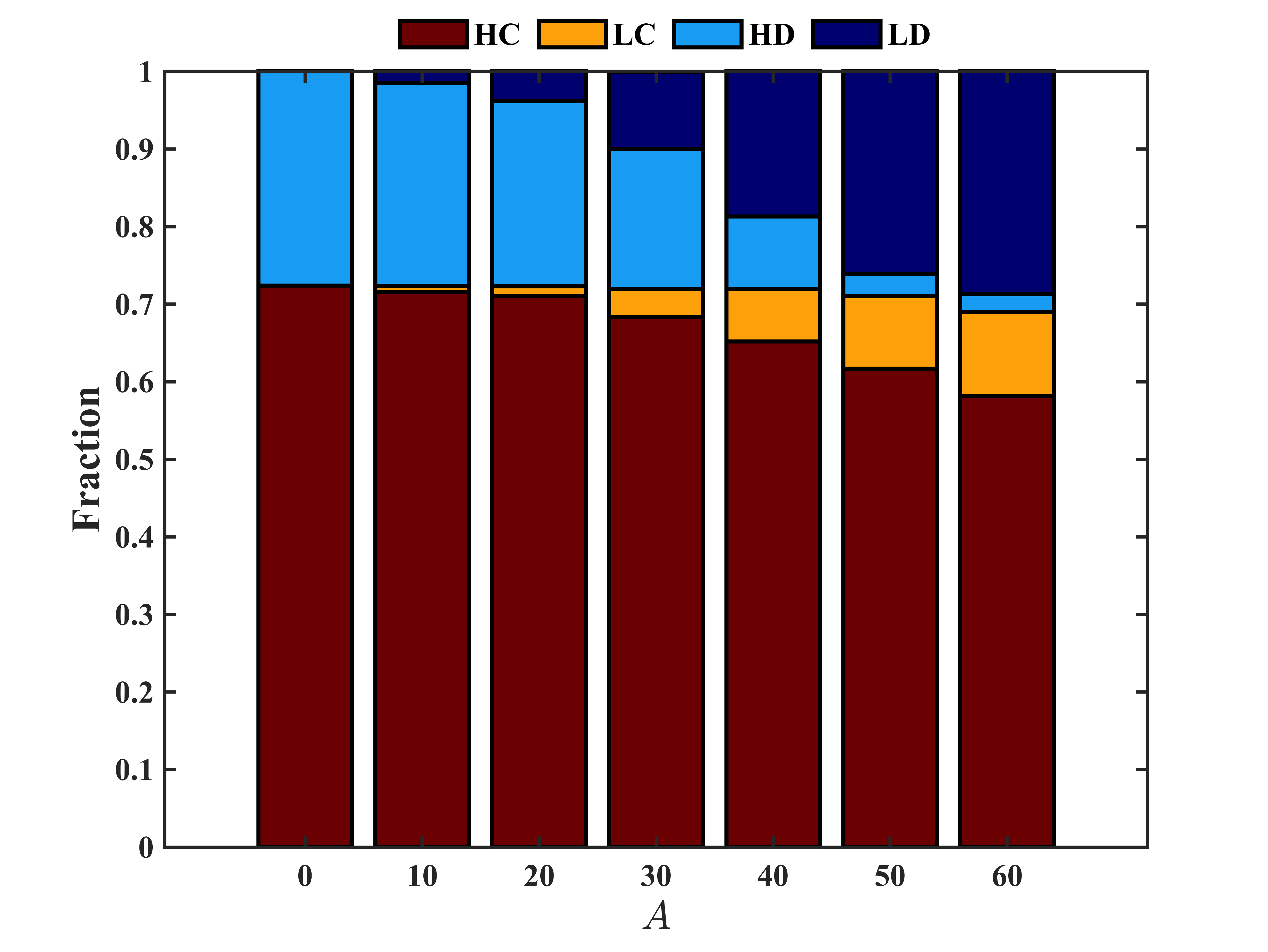}   
		\end{minipage}%
	}
	\subfloat[] 
	{
		\begin{minipage}[t]{0.3\textwidth}
			\centering      
			\includegraphics[width=1\textwidth]{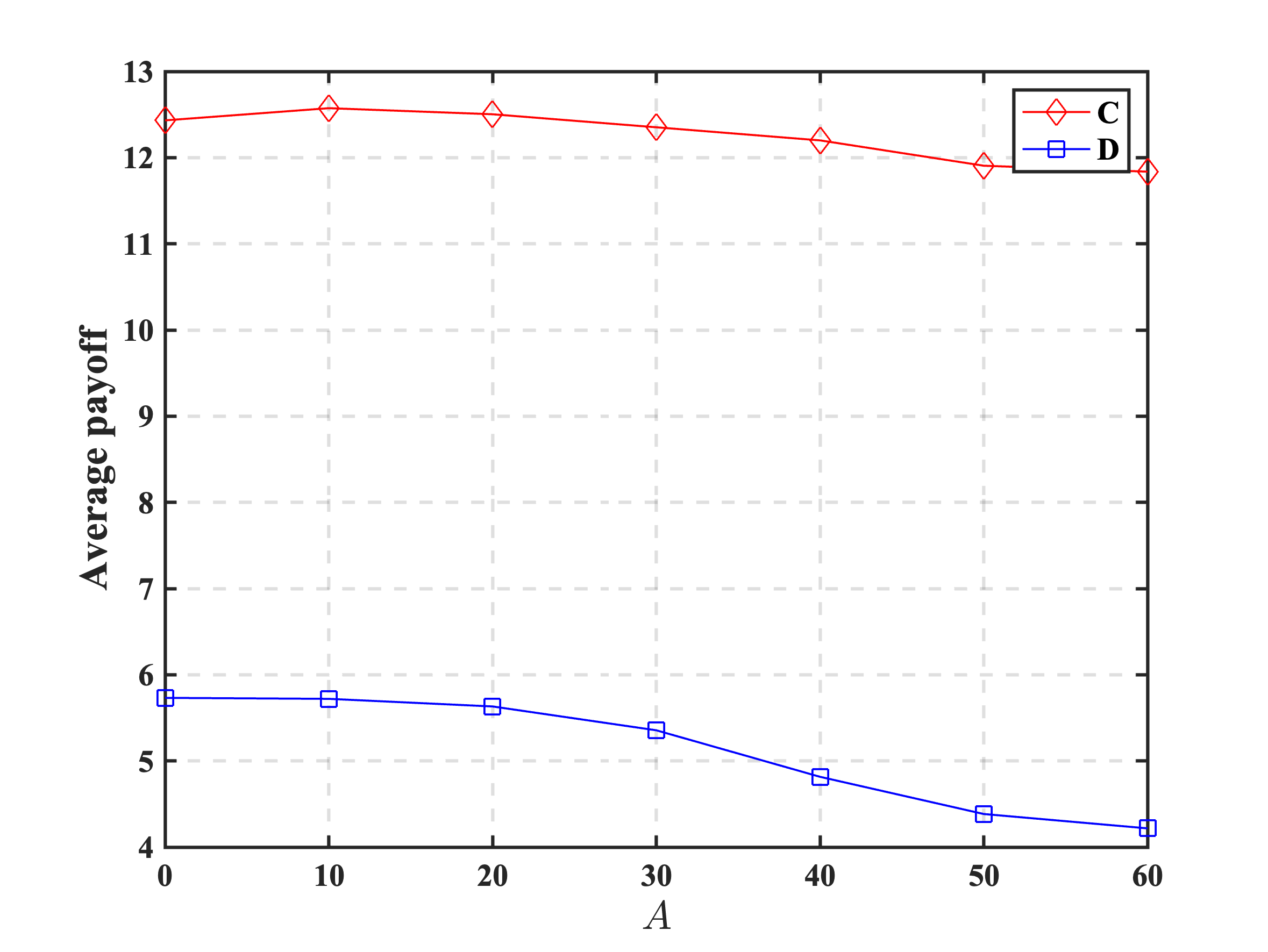}   
		\end{minipage}
	}
	\subfloat[] 
	{
		\begin{minipage}[t]{0.3\textwidth}
			\centering      
			\includegraphics[width=1\textwidth]{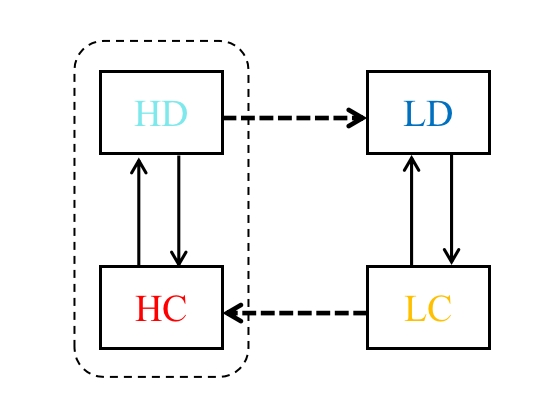}   
		\end{minipage}
	}%
        \\
        \subfloat[] 
	{
		\begin{minipage}[t]{0.3\textwidth}
			\centering          
			\includegraphics[width=1\textwidth]{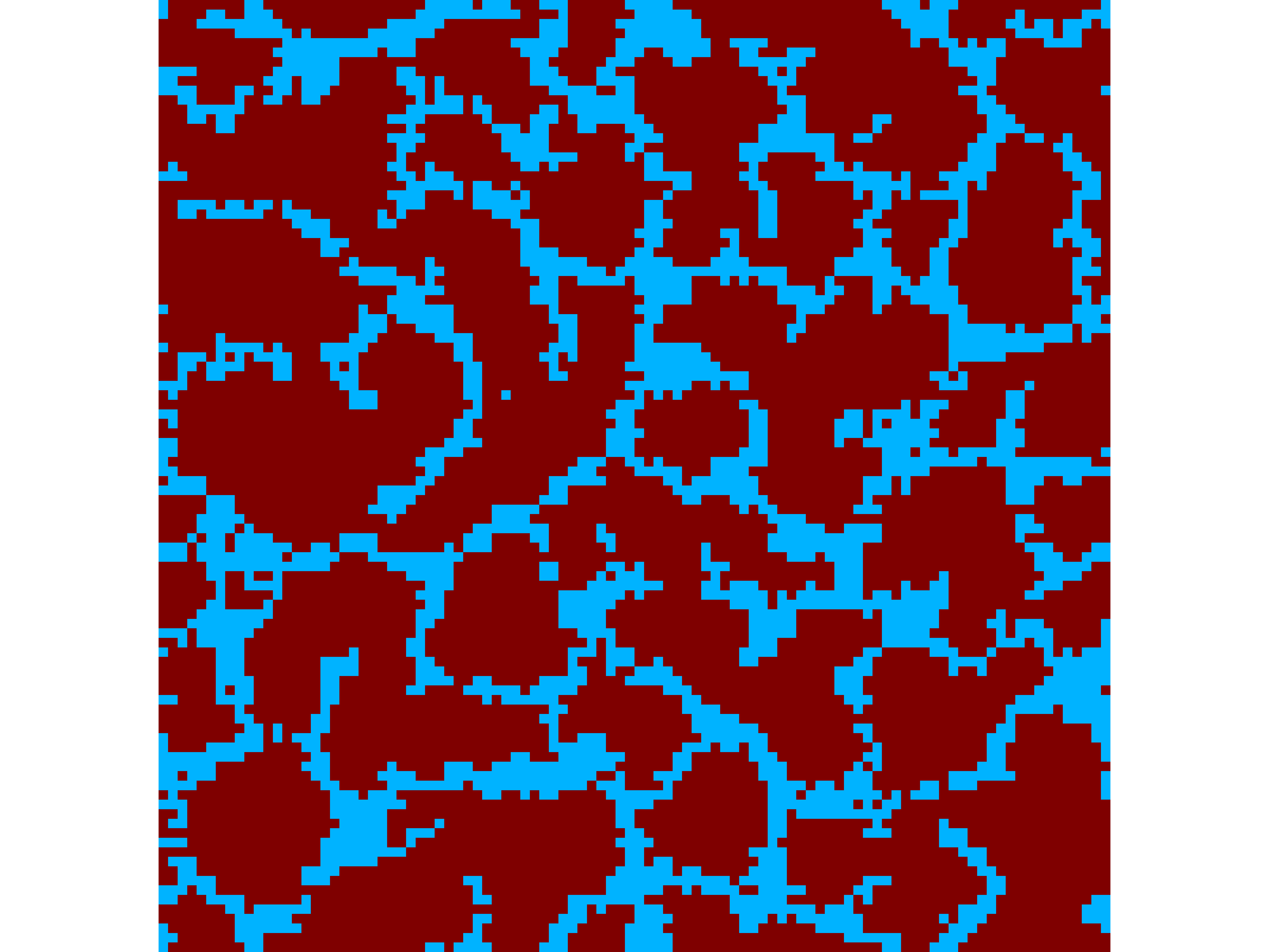}   
		\end{minipage}%
	}
	\subfloat[] 
	{
		\begin{minipage}[t]{0.3\textwidth}
			\centering      
			\includegraphics[width=1\textwidth]{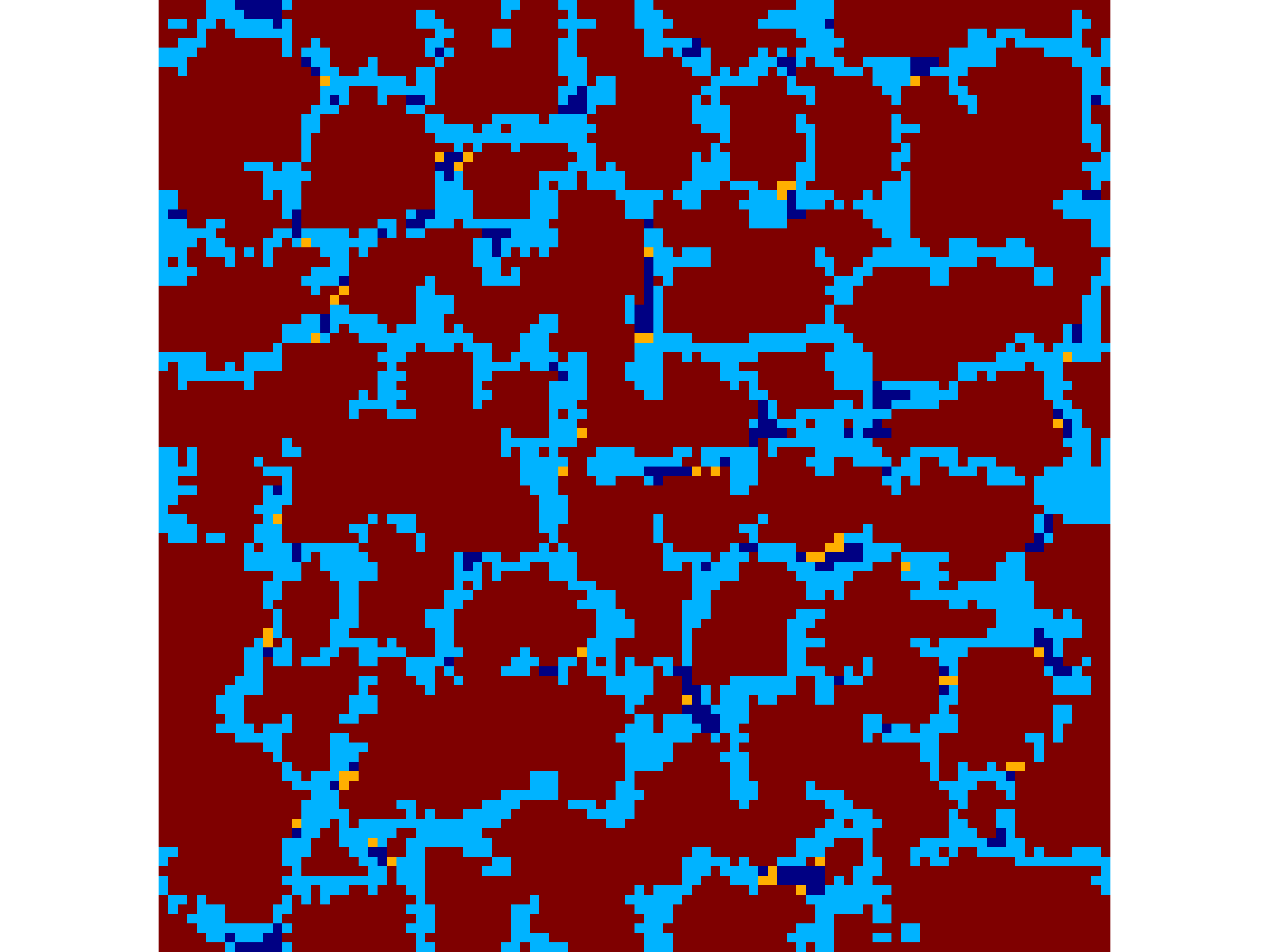}   
		\end{minipage}
	}
	\subfloat[] 
	{
		\begin{minipage}[t]{0.3\textwidth}
			\centering      
			\includegraphics[width=1\textwidth]{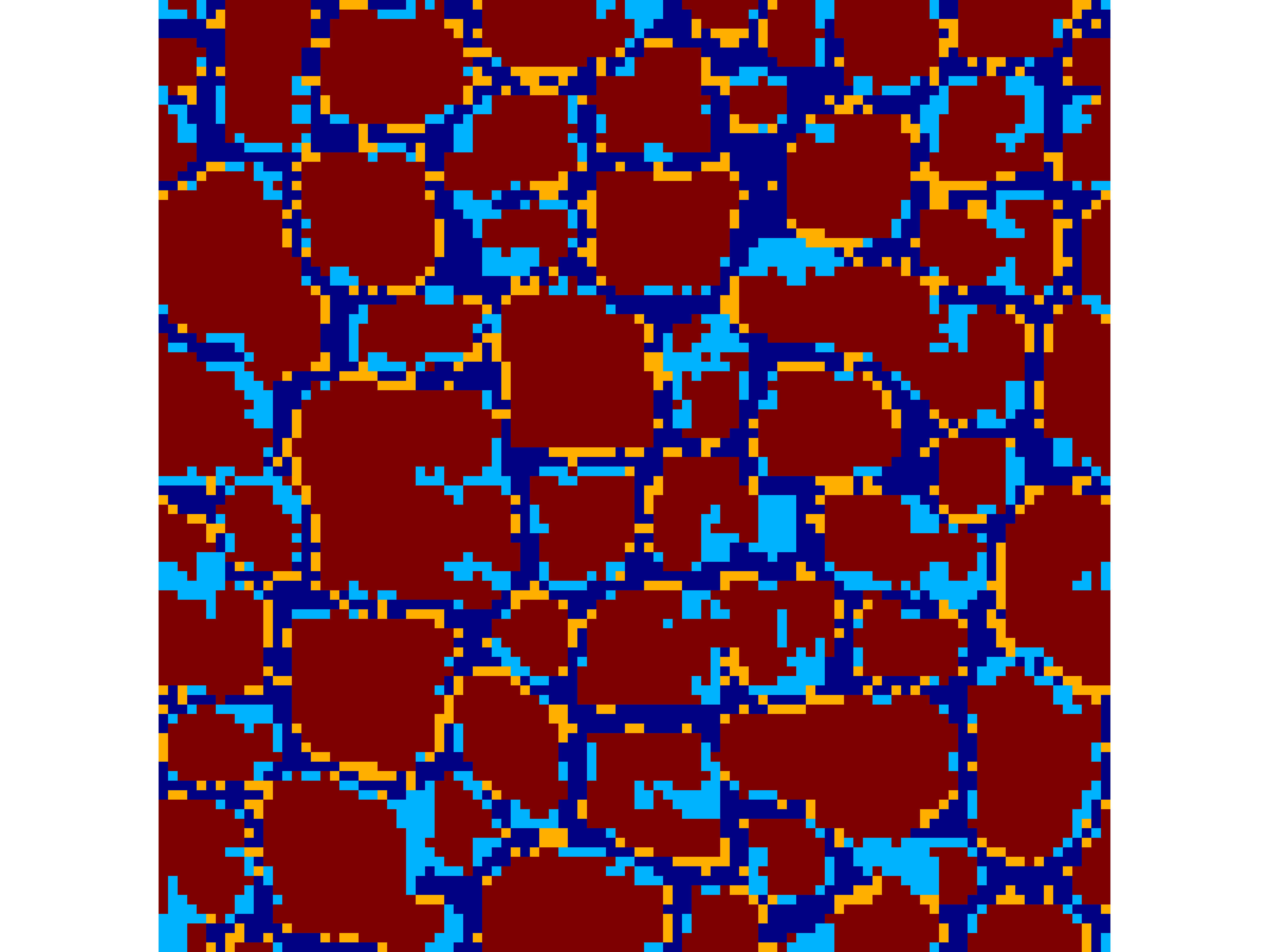}   
		\end{minipage}
	}%
	\caption{When \( A \in \{0, 10, 20, 30, 40, 50, 60\} \) and the population is in a evolutionary stable state, panel (a) shows the stacked chart of the fraction of the four types of groups, and panel (b) shows the average payoff for cooperators and defectors. Panel (c) shows the he transformation relationships for the four types of groups. Panel (d)-(f) show the snapshot when the population reaches stability when \(A = 0\), \(A = 10\) and \(A = 30\), respectively. In panel (a), (d)-(f), the color settings are the same as in Figure~\ref{fig_snap}. The fixed parameters are \( r = 3.3\) and \(\delta = 0.1 \).}\label{fig_discussion}
\end{figure}

Although the change in reputation threshold has little effect on the fraction of cooperators within certain intervals of \( r \) and \( \delta \) values, it can affect the weight of the four types of groups, the average payoffs of cooperators and defectors, and the population structure. In the stacked chart shown in Figure~\ref{fig_discussion}(a), as \( A \) increases, the overall fraction of cooperators (HC + LC) does not change significantly, but the proportions of the four types of groups do show different trends. When \( A = 0 \), only the synergy effect is active, and all groups in the population are synergy groups (HC + HD). When \( A > 0 \), both synergy and discounting effects can coexist in the population, and as \( A \) continues to increase, the number of synergy groups gradually decreases while the number of discounting groups (LC + LD) gradually increases. Interestingly, when \( A = 10 \) is relatively small, there are only a few discounting groups in the population, but the average payoff for cooperators is higher than when all groups are synergy groups, so the existence of a small number of discounting groups improves the payoffs for cooperators compared to when all are synergy groups. Furthermore, it can be noted that the increase in \( A \) has a greater impact on the payoffs of defectors; that is, increasing the reputation threshold will widen the gap between the average payoffs of cooperators and defectors.

The reason for the above phenomena can be explained by the relationship transformation diagram shown in Figure~\ref{fig_discussion}(c) and the snapshot diagrams shown in Figures~\ref{fig_discussion}(d)-(e). When \( A = 0 \), there is only mutual transformation between HD and HC (as shown by the dashed box in Figure~\ref{fig_discussion}(c)), and all players maintain a relatively high level of payoffs. Since player reputation evolves and according to Eq.~(\ref{eq_Ri}), the transformation of the four types of groups maintains a complete cycle when \( A > 0 \). The transformation processes indicated by the dashed arrows in Figure~\ref{fig_discussion}(c) are mainly controlled by the parameter \( A \). When \( A \) is relatively small, groups are easily defined as synergy groups, the transformation from HD to LD is difficult, and from LC to HC is easy. As \( A \) increases, the situation becomes the opposite. Comparing Figures~\ref{fig_discussion}(d) and (e), increasing \( A \) does not change the population structure, but compared to when all groups are synergy groups, these partially transformed LC groups bring extra payoffs to the group when participating in the PGG of HC, thereby slightly increasing the average payoffs of cooperators. Meanwhile, due to the presence of a small number of LD groups, the payoffs of defectors slightly decrease. As \( A \) further increases, the transformation from HD to LD becomes frequent, leading to an increase in the number of LD and LC groups, thereby reducing the average earnings of both cooperators and defectors. Furthermore, it can be seen in Figure~\ref{fig_discussion}(f) that clusters of cooperators mostly exist in the form of HC, and there is a large amount of LC at the boundary between defector and cooperator clusters. This has a greater impact on the payoffs of defectors, thus increasing the reputation threshold will widen the gap between the average payoffs of cooperators and defectors.

\section{Conclusion}
Indirect reciprocity is an important mechanism to promote cooperation among selfish individuals, and reputation as its direct expression has received widespread attention. Most previous studies have focused on the role of reputation in linear public goods games, but its impact on the numerous nonlinear public goods games that exist in the real world has not yet been deeply studied. In this paper, we explore the impact of reputation on the evolution of cooperation in spatial nonlinear public goods games, assuming that the reputation of all players within a group determine the group's overall reputation, which is defined as the average reputation of all players in the group. The reputation type of group is measured by a globally uniform reputation threshold, reflecting the policy and interest pattern of the society. Specifically, if the group's reputation is not lower than the reputation threshold, the group is defined as a high-reputation group, and vice versa for low-reputation groups. The payoff calculation for the high-reputation group reflects the synergy effect, while for the low-reputation group it reflects the discounting effect, where the synergy and discount effects are captured by a unified reputation impact factor.

By modeling on a square lattice, we find that reputation-based synergy and discounting mechanisms promote cooperation compared to the linear PGG, and the larger the reputation impact factor, the higher the fraction of cooperators. In addition, increasing the reputation impact factor can accelerate the phase transition of the population from full defection to full cooperation. When the reputation influence factor reaches a certain value, it is difficult for cooperators and defectors to coexist in the population. Notably, different enhancement factors and reputation impact factors can exhibit two states as the reputation threshold increases: when either is small, an increasing reputation threshold reduces the fraction of cooperators; but as they increase, the rise in the reputation threshold has a minimal impact on the fraction of cooperators in the population. Therefore, increasing the enhancement factor and reputation impact factor will, to some extent, weaken the influence of the reputation threshold on the population. Although increasing the reputation threshold does not change the overall fraction of cooperators within a certain range, it does affect the ratio of synergy groups to discounting groups, the average payoffs of cooperators and defectors, and the population structure. An increase in the reputation threshold reduces the number of synergy groups and increases the number of discounting groups. Interestingly, when the reputation threshold is low, the presence of a small number of discounting groups increases the average payoffs of cooperators. Increasing the reputation threshold decreases overall average payoffs but widens the gap between the average payoffs of cooperators and defectors.

Our work explores the evolution of cooperation in nonlinear PGG from a reputation perspective, and there is much future work to explore. For example, in this paper, simulations were performed only on square lattices, but the connections between players in the real world are complex, and one of the future works is to explore the impact of this reputation mechanism on more complex networks. Additionally, our work uses a unified reputation threshold and a unified reputation impact factor, and integrating the two parameters with the gaming environment is a perspective worth investigating. Overall, the reputation-based synergy and discounting mechanism proposed in this study can provide valuable insights for effectively solving dilemmas in the real world and may stimulate further related research.

\ack{This work is supported by National Key R$\&$D Program of China (2022ZD0116800), Program of National Natural Science Foundation of China (62141605, 12201026, 11922102, 11871004), and Beijing Natural Science Foundation (Z230001).}

\section*{Reference}
\bibliography{main.bib}

\end{document}